\documentclass[twocolumn]{jpsj3}

\usepackage{color}
\usepackage{bm}

\title{%
Disorder Effect on Chiral Edge Modes and Anomalous Hall Conductance
in Weyl Semimetals
}

\author{%
Yositake Takane
}

\inst{%
Department of Quantum Matter, Graduate School of Advanced Sciences of Matter,\\
Hiroshima University, Higashihiroshima, Hiroshima 739-8530, Japan
}

\recdate{ \hspace{50mm} }

\abst{%
Typical Weyl semimetals host chiral surface states and
hence show an anomalous Hall response.
Although a Weyl semimetal phase is known to be robust against weak disorder,
the effect of disorder on chiral states has not been fully clarified so far.
We study the behavior of such chiral states in the presence of disorder
and its consequences on an anomalous Hall response, focusing on a thin slab
of Weyl semimetal with chiral surface states along its edge.
It is shown that weak disorder does not disrupt chiral edge states
but crucially affects them owing to the renormalization of a mass parameter:
the number of chiral edge states changes depending on the strength of disorder.
It is also shown that the Hall conductance is quantized
when the Fermi level is located near Weyl nodes within a finite-size gap.
This quantization of the Hall conductance collapses
once the strength of disorder exceeds a critical value,
suggesting that it serves as a probe to distinguish a Weyl semimetal phase
from a diffusive anomalous Hall metal phase.
}

\begin{document}
\sloppy
\maketitle

\section{Introduction}

Weyl semimetals are three-dimensional (3D) topological systems
possessing pairs of nondegenerate Dirac cones
with opposite chirality.~\cite{volovik,murakami,wan,yang,burkov1,burkov2,WK,
delplace,halasz,sekine}
The band touching point of each Dirac cone is called a Weyl node.
A typical feature of Weyl semimetals is that low-energy states
with chirality appear on their surface~\cite{wan} under the condition
that a pair of Weyl nodes is separated in reciprocal space.
This is realized in the absence of time-reversal symmetry.
As these surface states connect a pair of Weyl nodes in reciprocal space,
they are called Fermi arc states.
The presence of Fermi arc states gives rise to
an anomalous Hall effect.~\cite{burkov1}
If a pair of Weyl nodes is separated in energy space as a consequence of
the breaking of inversion symmetry, the chiral magnetic effect,
another unusual electromagnetic response of Weyl semimetals, is expected to
appear.~\cite{nielsen,fukushima,zyuzin,son1,son2,basar,liu,vazifeh,chen1,
goswami1,landsteiner,chang,takane,sumiyoshi,fujita}
Thus far, TaAs and NbAs have experimentally been identified
as Weyl semimetals.~\cite{weng,huang1,xu1,lv1,lv2,xu2}

For Weyl semimetals as well as related Dirac semimetals, the effect of
disorder has been a subject of intensive study.~\cite{fradkin1,fradkin2,
shindou,goswami2,huang2,kobayashi,ominato,pixley,sbierski,biswas,nandkishore,
roy,chen2,shapourian,LOS,bera,gorbar,yoshimura}
It has been shown that a Weyl semimetal phase is robust against weak disorder;
it persists up to some finite critical strength of disorder,
at which a transition to a diffusive anomalous Hall metal phase occurs.
The boundary between the two distinct phases has been determined under
the criterion that, at the boundary, the localization length becomes
scale-invariant.~\cite{chen2,shapourian,LOS}
However, the meaning of the phase boundary is slightly unclear in the sense
that both phases are metallic and have a finite Hall conductivity.
To find a clear difference between them, we need to consider the density of
bulk states at Weyl nodes, which jumps to a finite value from zero
at the transition to a diffusive metal phase.~\cite{fradkin2,shindou,
goswami2,huang2,kobayashi,ominato,pixley,LOS}
Note that, in previous studies on this subject, the role of Fermi arc
surface states is not explicitly considered,
except in Refs.~\citen{LOS}, \citen{gorbar}, and \citen{yoshimura}.
As Fermi arc states are expected to be stabilized only in
a Weyl semimetal phase, it is meaningful to study their response
against disorder to uncover a clear difference between the two phases.

For definiteness, we focus on a prototypical system of Weyl semimetals
with a pair of Weyl nodes
at $\mib{k}_{\pm} = (0,0,\pm k_{0})$ in the 3D Brillouin zone.
If $\mib{k}_{\pm}$ are projected onto $\tilde{\mib k}_{\pm}$
in the surface Brillouin zone corresponding to a particular flat surface,
Fermi arc surface states appear in a manner to connect
$\tilde{\mib k}_{+}$ and $\tilde{\mib k}_{-}$.
This means that, on a flat surface parallel to the $xy$-plane,
Fermi arc states disappear as the projected points coincide
at $(k_{x},k_{y}) = (0,0)$ in the surface Brillouin zone.
If the system is in the shape of a slab
with its top and bottom surfaces being parallel to the $xy$-plane,
Fermi arc states appear only on its side.~\cite{okugawa}
An important feature of these surface states is that they are chiral;
they propagate only in a given direction circulating the system along its edge.
That is, they are similar to chiral edge states in an ordinary quantum Hall
system under a strong magnetic field.~\cite{klitzing,buttiker}
This setup is suitable to examine the disorder effect on chiral edge states
and an anomalous Hall response related to them.

In this paper, we study the disorder effect on Weyl semimetals
taking the role of surface states into explicit consideration.
Our attention is focused on a thin slab of Weyl semimetal
hosting chiral edge modes only at its side.
Setting it in a Hall bar geometry with source, drain, and
voltage electrodes, we numerically calculate
the dimensionless Hall conductance $G_{\rm H}$ in the presence of disorder.
It is shown that, when the Fermi level is located at the Weyl nodes,
$G_{\rm H}$ is quantized to an integer equal to the number of chiral modes.
This quantization of $G_{\rm H}$ is considered to be stabilized by
a finite-size gap for bulk states at the Weyl nodes; thus, it collapses once
the strength $W$ of disorder exceeds the critical value $W_{\rm c}$ and
hence the density of bulk states at the Weyl nodes becomes finite.
This indicates that a Weyl semimetal phase can be clearly distinguished
from a diffusive anomalous Hall metal phase by the presence or absence of
the quantization of $G_{\rm H}$.
It is also shown that the quantized value of $G_{\rm H}$
increases with increasing $W$ as long as $W < W_{\rm c}$.
This reflects the increase in the number of chiral modes
caused by the renormalization of a mass parameter due to disorder,
and suggests that the property of chiral modes can be controlled by disorder.

In the next section, we present a tight-binding model for Weyl semimetals
and analyze the behavior of chiral edge modes.
Particularly, we demonstrate how many chiral edge modes are allowed
for a given set of parameters in a slab of Weyl semimetal.
In Sect.~3, we numerically study the dimensionless Hall conductance
and related transport coefficients in a Hall bar geometry.
The last section is devoted for summary and discussion.
We set $\hbar = 1$ throughout this paper.

\section{Model}

For Weyl semimetals with a pair of Weyl nodes
at $\mib{k}_{\pm} = (0,0,\pm k_{0})$, we introduce a tight-binding model
on a cubic lattice with the lattice constant $a$.
The indices $l$, $m$, and $n$ are respectively used to specify lattice sites
in the $x$-, $y$-, and $z$-directions,
and the two-component state vector for the $(l,m,n)$th site is expressed as
\begin{align}
  |l,m,n \rangle
  =  \left[ |l,m,n \rangle_{\uparrow}, |l,m,n \rangle_{\downarrow} \right] ,
\end{align}
where $\uparrow, \downarrow$ represents the spin degree of freedom.
The tight-binding Hamiltonian is given by
$H = H_{0}+H_{x}+H_{y}+H_{z}$ with~\cite{yang,burkov1,imura1}
\begin{align}
   H_{0}
 & = \sum_{l,m,n} |l,m,n \rangle h_{0} \langle l,m,n| ,
         \\
   H_{x}
 & = \sum_{l,m,n}
     \left\{ |l+1,m,n \rangle h_{x}^{+} \langle l,m,n|
             + {\rm h.c.} \right\} ,
         \\
   H_{y}
 & = \sum_{l,m,n}
     \left\{ |l,m+1,n \rangle h_{y}^{+} \langle l,m,n|
             + {\rm h.c.} \right\} ,
         \\
   H_{z}
 & = \sum_{l,m,n}
     \left\{ |l,m,n+1 \rangle h_{z}^{+} \langle l,m,n|
             + {\rm h.c.} \right\} .
\end{align}
Here, the $2 \times 2$ matrices are
\begin{align}
   h_{0}
 & = \left[ 
       \begin{array}{cc}
         2t\cos(k_{0}a) + 4B & 0 \\
         0 & -2t\cos(k_{0}a) - 4B
       \end{array}
     \right] ,
               \\
   h_{x}^{+}
 & = \left[ 
       \begin{array}{cc}
         -B & \frac{i}{2}A \\
         \frac{i}{2}A & B
       \end{array}
     \right] ,
               \\
   h_{y}^{+}
 & = \left[ 
       \begin{array}{cc}
         -B & \frac{1}{2}A \\
         -\frac{1}{2}A & B
       \end{array}
     \right] ,
               \\
   h_{z}^{+}
 & = \left[ 
       \begin{array}{cc}
         -t & 0 \\
         0 & t
       \end{array}
     \right] ,
\end{align}
where $\pi/a > k_{0} > 0$, and the other parameters, $A$, $B$, and $t$,
are assumed to be real and positive.
The Fourier transform of $H$ is expressed as
$\mathcal{H}(k_{x},k_{y},k_{z}) = \mathcal{H}_{x}(k_{x})
+\mathcal{H}_{y}(k_{y})+\mathcal{H}_{z}(k_{z})$ with
their representation matrices given by
\begin{align}
   h_{x}(k_{x})
 & = \left[ 
        \begin{array}{cc}
          2B[1-\cos(k_{x}a)] & A\sin(k_{x}a) \\
          A \sin(k_{x}a) & -2B[1-\cos(k_{x}a)]
        \end{array}
     \right] ,
            \\
  h_{y}(k_{y})
 & = \left[ 
       \begin{array}{cc}
         2B[1-\cos(k_{y}a)] & -i A\sin(k_{y}a) \\
         i A\sin(k_{y}a) & -2B[1-\cos(k_{y}a)]
       \end{array}
     \right] ,
            \\
  h_{z}(k_{z})
 & = \left[ 
       \begin{array}{cc}
         \Delta(k_{z}) & 0 \\
         0 & -\Delta(k_{z})
       \end{array}
     \right] ,
\end{align}
where
\begin{align}
  \Delta(k_{z}) = - 2t\left[\cos(k_{z}a)-\cos(k_{0}a)\right] .
\end{align}

From the expression of $\mathcal{H}(k_{x},k_{y},k_{z})$,
we find that the energy dispersion of this model is
\begin{align}
           \label{eq:exp-E}
   E =
 & \pm \Big\{\left[\Delta(k_{z})
                   +2B\bigl(2-\cos(k_{x}a)-\cos(k_{y}a)\bigr)\right]^{2}
              \nonumber \\
 & \hspace{10mm}
           + A^{2}\bigl(\sin^{2}(k_{x}a)+\sin^{2}(k_{y}a)\bigr)
       \Big\}^{\frac{1}{2}} .
\end{align}
Equation~(\ref{eq:exp-E}) indicates that a pair of Weyl nodes appears
only at $\mib{k}_{\pm} = (0,0,\pm k_{0})$ under the assumption of
\begin{align}
      \label{eq:assump_B-t}
  2B > t \bigl(1-\cos(k_{0}a)\bigr) .
\end{align}
As noted in Sect.~1,
if the system is in the shape of a rectangular parallelepiped
with its top and bottom surfaces being parallel to the $xy$-plane,
Fermi arc states appear only on the side surfaces
in the form of chiral edge modes.

\begin{figure}[tbp]
\begin{center}
\includegraphics[height=2.5cm]{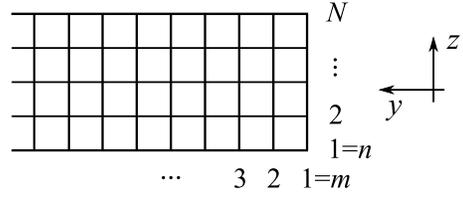}
\end{center}
\caption{
Cross section of the system considered in the text;
it consists of $N$ sites in the $z$-direction and
is semi-infinite in the $y$-direction.
}
\end{figure}
Let us consider chiral edge modes in the system of thickness $N$
stacked in the $z$-direction, focusing on the case where
the system has a flat surface parallel to the $xz$-plane.
Accordingly, we assume that the system occupies the region of
$N \ge n \ge 1$ in the $z$-direction and that of $m \ge 1$
in the $y$-direction (see Fig.~1), and is infinitely long in the $x$-direction.
Thus, the system has an infinitely long surface of width $N$
parallel to the $xz$-plane at $m = 1$.
In this setup, we consider low-energy states localized near the surface.

To begin with, we examine the case
where the periodic boundary condition is imposed in the $z$-direction.
Although this condition is rather artificial, the resulting argument
provides us a concrete basis to describe actual situations.
Since the system is assumed to be infinite in the $x$-direction,
the wave number $k_{x}$ in the $x$-direction becomes a good quantum number.
Furthermore, since the periodic boundary condition is imposed
in the $z$-direction, the $z$-component of a wave function can be
characterized by the wave number $k_{z}$ in the $z$-direction.
With these observations, we express eigenstates of the system
in the following form:
\begin{align}
        \label{eq:psi_kx-kz}
    |\psi(k_{x},k_{z}) \rangle
    = \sum_{m=1}^{\infty}
     |k_{x},m,k_{z} \rangle \mib{\varphi}(m) ,
\end{align}
where $|k_{x},m,k_{z}\rangle$ denotes the partial Fourier transform of
$|l,m,n\rangle$ with respect to $l$ and $n$, and $\mib{\varphi}(m)$ is
the two-component wave function in the $y$-direction.
In accordance with $|k_{x},m,k_{z}\rangle$, we introduce
the partial Fourier transform of $H$ with respect to $l$ and $n$,
which is defined by $\mathcal{H}(k_{x},k_{z}) = \tilde{H}_{0}
+ \mathcal{H}_{x}(k_{x}) + H_{y} + \mathcal{H}_{z}(k_{z})$,
where the matrix representation of $\tilde{H}_{0}$ on the basis of
$\{ |k_{x},m,k_{z}\rangle \}$ is
\begin{align}
   \tilde{h}_{0}(k_{z})
   = \left[ 
        \begin{array}{cc}
          2B & 0 \\
          0 & - 2B
        \end{array}
     \right] .
\end{align}
We now approximately obtain low-energy eigenstates of
$\mathcal{H}(k_{x},k_{z})$ localized near the surface at $m = 1$.
The procedure is similar to that of Ref.~\citen{arita} used to derive
an effective Hamiltonian for surface states of weak topological insulators.
Let us solve the eigenvalue equation for the $y$-direction:
\begin{align}
      \label{eq:EE-x}
   H_{\perp}(k_{z})|\psi(k_{x},k_{z}) \rangle
   = E_{\perp}|\psi(k_{x},k_{z}) \rangle ,
\end{align}
where
\begin{align}
   H_{\perp}(k_{z}) \equiv \tilde{H}_{0}+H_{y}+\mathcal{H}_{z}(k_{z}) .
\end{align}
Since its solutions localized near the surface are necessary for our argument,
the appropriate boundary condition for $\mib{\varphi}(m)$ is
$\mib{\varphi}(0) = \mib{\varphi}(\infty) = {}^{t}(0,0)$.
Solving Eq.~(\ref{eq:EE-x}) under the required boundary condition,
we obtain the solution with $E_{\perp} = 0$ (see Appendix) as
\begin{align}
     \label{eq:varphi_0}
   \mib{\varphi}(m)
   = \mathcal{C} \left(\rho_{+}^{m}-\rho_{-}^{m}\right)
     \frac{1}{\sqrt{2}}
     \left[ \begin{array}{c}
              1 \\
              1
            \end{array}
     \right] ,
\end{align}
where $\mathcal{C}$ is a normalization constant, and
$\rho_{+}$ and $\rho_{-}$ are constants given by Eq.~(\ref{eq:rho-pm})
satisfying $|\rho_{\pm}| < 1$.
Clearly, $\mathcal{C}(\rho_{+}^{m}-\rho_{-}^{m})$ represents the penetration of
surface states into the bulk.
Note that $|\psi(k_{x},k_{z}) \rangle$ becomes an exact eigenstate of
$\mathcal{H}(k_{x},k_{z})$ at $k_{x} = 0$ as $\mathcal{H}_{x}(k_{x})$ vanishes.
From this fact, we can regard $|\psi(k_{x},k_{z})\rangle$ as
an approximate eigenstate of $\mathcal{H}(k_{x},k_{z})$
for small $k_{x}$ satisfying $|k_{x}|a \ll \pi/2$.
By treating $\mathcal{H}_{x}(k_{x})$ as a perturbation,
the corresponding eigenvalue of energy is expressed as
\begin{align}
 E = \langle\psi(k_{x},k_{z})|\mathcal{H}_{x}(k_{x})|\psi(k_{x},k_{z})\rangle ,
\end{align}
which results in
\begin{align}
   E = A \sin(k_{x}a)
\end{align}
in the regime of $|k_{x}|a \ll \pi/2$.
As this mode has a linear energy dispersion with a positive velocity, we should
identify it as a chiral edge mode propagating in the positive $x$-direction.
Note that, as shown in Appendix, $|\rho_{\pm}| < 1$ holds for $k_{z}$
within the interval of $k_{0} > k_{z} > -k_{0}$.
Thus, a chiral edge mode appears for each $k_{z}$ satisfying
\begin{align}
       \label{eq:interval-k_z}
  k_{0} > k_{z} > -k_{0} .
\end{align}
This is consistent with a heuristic argument that Fermi arc surface states
appear in a manner to connect $\tilde{\mib k}_{\pm}$,
where $\tilde{\mib k}_{\pm} = (0,\pm k_{0})$ on the $k_{x}k_{z}$-plane
in this case.

We then turn to the case with the open boundary condition,
which is appropriate for actual systems with a finite thickness.
It is convenient to rewrite Eq.~(\ref{eq:psi_kx-kz}) as
\begin{align}
        \label{eq:psi_kx}
    |\psi(k_{x},k_{z}) \rangle
    = \sum_{m=1}^{\infty}\sum_{n=1}^{N}
     |k_{x},m,n \rangle \mib{\phi}(m,n;k_{z})
\end{align}
with
\begin{align}
   \mib{\phi}(m,n;k_{z})
   = \frac{1}{\sqrt{N}}e^{ik_{z}an}\mib{\varphi}(m) .
\end{align}
Here, $|k_{x},m,n\rangle$ denotes the partial Fourier transform of
$|l,m,n\rangle$ with respect to $l$.
To construct a wave function that satisfies the open boundary condition,
we superpose a pair of eigenstates with $k_{z}$ and $-k_{z}$:
\begin{align}
    |\Psi(k_{x},k_{z}) \rangle
    = \sum_{m=1}^{\infty}\sum_{n=1}^{N}
     |k_{x},m,n \rangle \mib{\Phi}(m,n;k_{z})
\end{align}
with
\begin{align}
        \label{eq:Psi-pair}
  \mib{\Phi}(m,n;k_{z})
  = \frac{1}{\sqrt{2}}
    \bigl( \mib{\phi}(m,n;k_{z}) - \mib{\phi}(m,n;-k_{z}) \bigr) ,
\end{align}
where $\pi/a > k_{z} > 0$.
Note that the open boundary condition in the $z$-direction is satisfied
when $\mib{\Phi}(m,0;k_{z}) = \mib{\Phi}(m,N+1;k_{z}) =  {}^{t}(0,0)$.
Clearly, $\mib{\Phi}(m,n;k_{z})$ in Eq.~(\ref{eq:Psi-pair})
vanishes at $n = 0$ and hence
it satisfies the required boundary condition if it also vanishes at $n = N+1$.
This is the case when $k_{z}$ is given by
\begin{align}
  k_{z}a = \frac{j\pi}{N+1}
\end{align}
with $j = 1, 2, \dots, N$.
Combining this with Eq.~(\ref{eq:interval-k_z}),
we find that a chiral edge mode appears for each $j$ satisfying
\begin{align}
        \label{eq:condition-cm}
  k_{0}a > \frac{j\pi}{N+1} > 0 .
\end{align}
This equation determines the number of chiral edge modes stabilized
in the system of a finite thickness $N$.

\begin{figure}[tbp]
\begin{center}
\includegraphics[height=5.5cm]{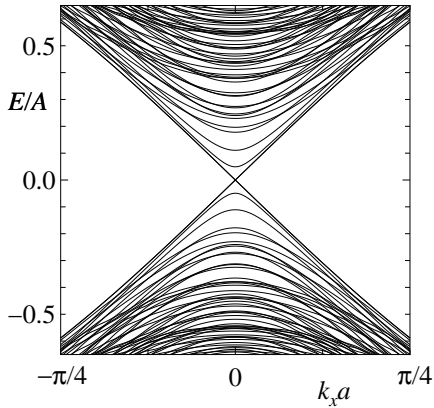}
\end{center}
\caption{
Band structure as a function of $k_{x}$
in the case of $N = 14$ and $k_{0}a = (6.9\pi)/15$ with $M = 45$.
}
\end{figure}
To examine the validity of the above argument,
we calculate the band structure of an infinitely long system of
cross-sectional area $M \times N$
as a function of the wave number $k_{x}$.
The above analysis indicates that chiral modes propagating
in the positive $x$-direction appear on the right edge (i.e., $m = 1$),
while those propagating in the negative $x$-direction appear
on the left edge (i.e., $m = M$).
Note that $N$ and $k_{0}$ directly affect the behavior of chiral modes
as is clear from Eq.~(\ref{eq:condition-cm}),
while $M$ is irrelevant for them as long as it is so large that
the coupling of counterpropagating chiral modes is negligible.
The result for the case of $N = 14$ and $k_{0}a = (6.9\pi)/15$
with $M = 45$ is shown in Fig.~2, where the other parameters are
$B/A = 0.5$ and $t/A = 0.5$.
We clearly observe the presence of the propagating modes
with a gapless linear energy dispersion.
Although it is unclear from this figure, the number of modes
in one propagating direction is equal to six and the right-going (left-going)
modes are localized near the right (left) edge.
Therefore, these modes should be identified  as the chiral edge modes.
Note that Eq.~(\ref{eq:condition-cm}) demonstrates that
the chiral edge modes are allowed for $6 \ge j \ge 1$ in this case.
This accounts for the number of modes
observed in the numerical calculation.

\section{Simulation of Electron Transport}

We study the electron transport in a disordered Weyl semimetal
at zero temperature by numerical simulation.
Particularly, our attention is focused on
the anomalous Hall effect due to chiral edge modes.
We consider the Weyl semimetal in a rectangular parallelepiped shape
of volume $L \times M \times N$,
to which six electrodes are attached to form a Hall bar geometry
as shown in Fig.~3.
Here, $N$ (not designated in Fig.~3) represents
the height of the system.
Planar electrodes 1 and 6 of area $M \times N$ respectively serve
as the source and drain electrodes,
and other line electrodes of width $N$ serve as voltage probes.
That is, the electric current $I$ is supplied from electrode 6 and
subtracted from electrode 1,
and no electric current flows through the other electrodes.

In analyzing the electron transport in this system,
the effect of bulk states should be taken into consideration.
Therefore, the scattering approach of B\"{u}ttiker~\cite{buttiker}
is inappropriate as it treats only chiral edge modes.
We thus employ a Green's function approach of Ref.~\citen{datta},
which is briefly described below.
With $\mib{r} \equiv (l,m,n)$ denoting the position of each site in the system,
$|l,m,n\rangle$ is rewritten as $|\mib{r}\rangle$.
Let us define Green's function as
\begin{align}
  G = \left( E_{\rm F}{\mib 1}-H-\Sigma \right)^{-1} ,
\end{align}
where ${\mib 1}=\sum_{\mib r}|\mib{r}\rangle\langle\mib{r}|$,
$E_{\rm F}$ is the Fermi energy, and $\Sigma$ is the self-energy
describing the coupling of the system with the electrodes.
We assume that $\Sigma$ is decomposed as $\Sigma = \sum_{p=1}^{6} \Sigma_{p}$,
and $\Sigma_{p}$, arising from the coupling with the $p$th electrode,
is given by
\begin{align}
  \Sigma_{p}
  =  -i\gamma \sum_{\mib{r} \in {\rm S}_{p}}
       |\mib{r}\rangle\langle\mib{r}| ,
\end{align}
where $\gamma$ is the coupling strength and ${\rm S}_{p}$ denotes
the set of sites in direct contact with the $p$th electrode.
In terms of Green's function, the transmission function
from the $q$th electrode to the $p$th electrode is defined as
\begin{align}
  T_{pq}
  = \Tr \left\{\Gamma_{p} G \Gamma_{q} G^{\dagger}\right\} ,
\end{align}
where $\Gamma_{p} \equiv i(\Sigma_{p} -\Sigma_{p}^{\dagger})$.
The electric current from the $q$th electrode to
the $p$th electrode is expressed as
\begin{align}
  I_{pq} = \frac{e}{2\pi}T_{pq}\left(\mu_{p}-\mu_{q}\right) ,
\end{align}
where $\mu_{p}$ is the chemical potential at the $p$th electrode.
Hence, the current $I_{p}$ flowing out from the $p$th electrode is written as
\begin{align}
   I_{p} = \frac{e}{2\pi}\sum_{q (\neq p)}
           T_{pq}\left(\mu_{p}-\mu_{q}\right) .
\end{align}
In accordance with the assumption given above,
the chemical potentials are determined by the following equations:
$I_{1} = -I_{6} = I$, and $I_{2} = I_{3} = I_{4} = I_{5} = 0$.
Here, we set $\mu_{6} = 0$ without loss of generality.
Owing to the current conservation of $\sum_{p=1}^{6}I_{p} = 0$,
we are allowed to omit one of the six equations.
Thus, we can obtain $\mu_{1}, \mu_{2}, \dots, \mu_{5}$ by simultaneously
solving the set of algebraic equations for $I_{1}, I_{2}, \dots, I_{5}$
once all the transmission functions are given.
The dimensionless Hall resistance between electrodes 2 and 3
and the dimensionless longitudinal resistance between electrodes 2 and 4
are respectively given by
\begin{align}
  R_{\rm H} & = \frac{e^{2}}{2\pi}\frac{\mu_{2}-\mu_{3}}{eI} ,
         \\
  R_{24} & = \frac{e^{2}}{2\pi}\frac{\mu_{2}-\mu_{4}}{eI} ,
\end{align}
and the dimensionless two-terminal resistance
between electrodes 1 and 6 is given by
\begin{align}
  R_{\rm SD} = \frac{e^{2}}{2\pi}\frac{\mu_{1}}{eI}
\end{align}
as $\mu_{6}$ is set equal to zero.
The dimensionless Hall conductance and dimensionless two-terminal
conductance are respectively defined as
\begin{align}
  G_{\rm H} & = R_{\rm H}^{-1} ,
        \\
  G_{\rm SD} & = R_{\rm SD}^{-1} .
\end{align}
\begin{figure}[tbp]
\begin{center}
\includegraphics[height=3.6cm]{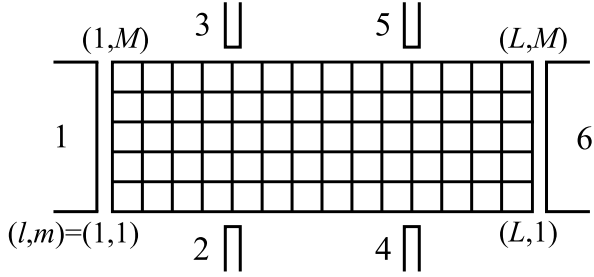}
\end{center}
\caption{
Top view of the lattice system with six electrodes, where the coordinates of
four corner sites are specified in the form of $(l,m)$.
The wide electrodes $1$ and $6$ are respectively in planar contact with
the sites on the left surface at $l = 1$ and
those on the right surface at $l = L$.
The thin electrodes $2$, $3$, $4$, and $5$ are in line contact with the sites
on the columns at $(l,m) = (L/3,1)$,
$(L/3,M)$, $(2L/3+1,1)$, and $(2L/3+1,M)$, respectively.
}
\end{figure}

We treat the system of $N = 14$, $M = 45$, and $L = 135$,
and use the following parameters: $k_{0}a = (6.9\pi)/15$, $B/A = 0.5$,
$t/A = 0.5$, and $\gamma/A = 0.4$.
Under these parameters, six chiral edge modes are stabilized
as in the case treated in the previous section.
The effect of disorder is incorporated by
adding the impurity potential term,
\begin{align}
  H_{\rm imp}
  = \sum_{\mib r} |\mib{r} \rangle
                    \left[ \begin{array}{cc}
                             V_{1}(\mib{r}) & 0 \\
                             0 & V_{2}(\mib{r})
                           \end{array} \right]
                  \langle \mib{r}| ,
\end{align}
to the Hamiltonian $H$.
We assume that $V_{1}$ and $V_{2}$ are uniformly distributed
within the interval of $[-W/2,+W/2]$.
That is, $W$ controls the strength of disorder.
We calculate the ensemble averages,
$\langle G_{\rm H}\rangle$, $\langle G_{\rm SD}\rangle$,
$\langle R_{\rm H}\rangle$, and $\langle R_{\rm 24}\rangle$,
over samples with different impurity configurations for a given value of $W/A$.
We also calculate the fluctuations of $G_{\rm H}$ and $G_{\rm SD}$ defined by
$\Delta G_{X} \equiv \sqrt{\langle G_{X}^{2}\rangle-\langle G_{X}\rangle^{2}}$,
where $X = {\rm H}$ or $\rm SD$.
In actual numerical calculations, $100$ samples are used to perform
the ensemble average at each data point.

\begin{figure}[tbp]
\begin{center}
\includegraphics[height=6.0cm]{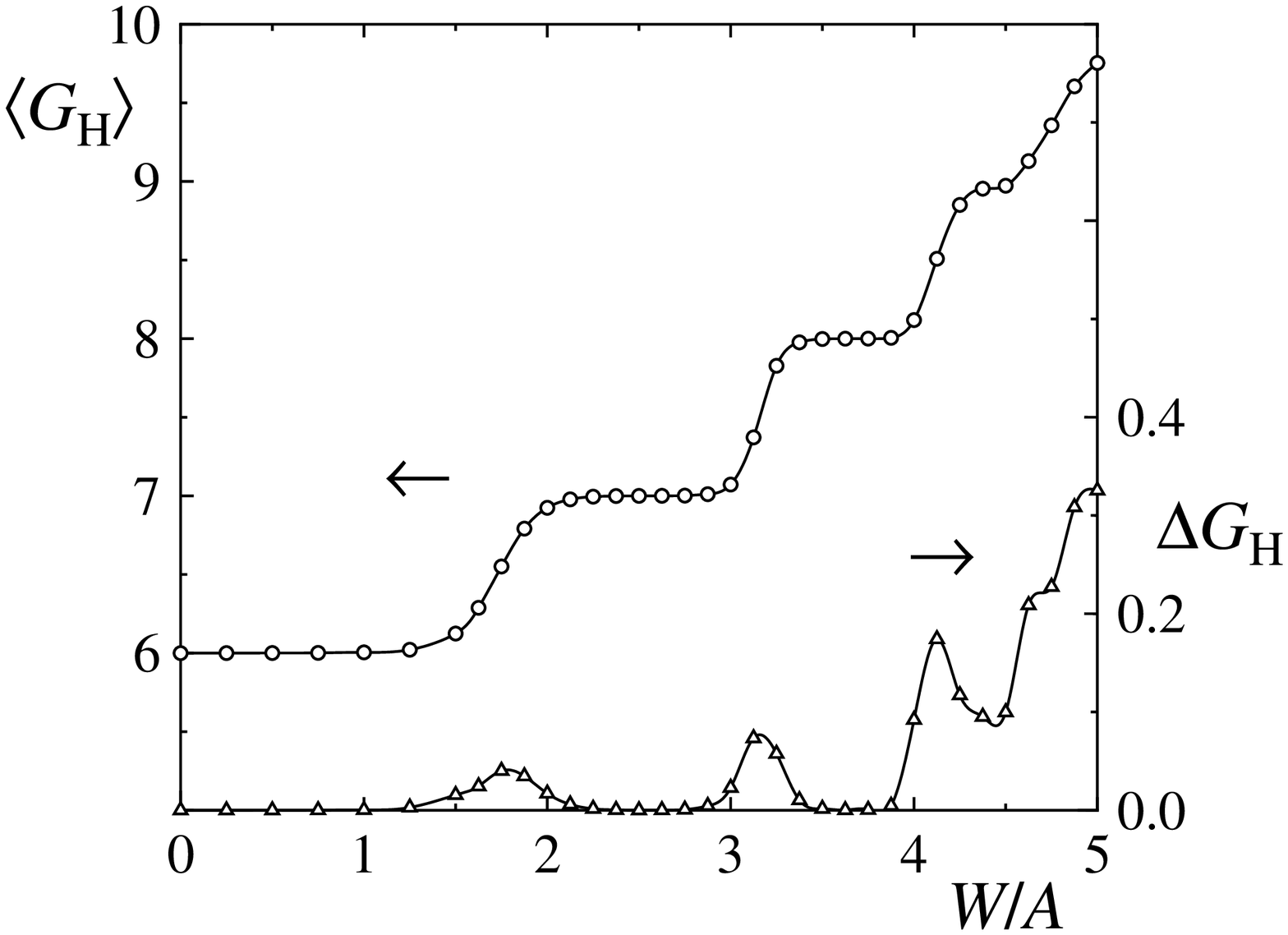}
\end{center}
\caption{
$\langle G_{\rm H} \rangle $ and $\Delta G_{\rm H}$
as functions of $W/A$ in the case of $E_{\rm F}/A = 0$.
Solid lines serve as visual guides.
}
\begin{center}
\includegraphics[height=6.0cm]{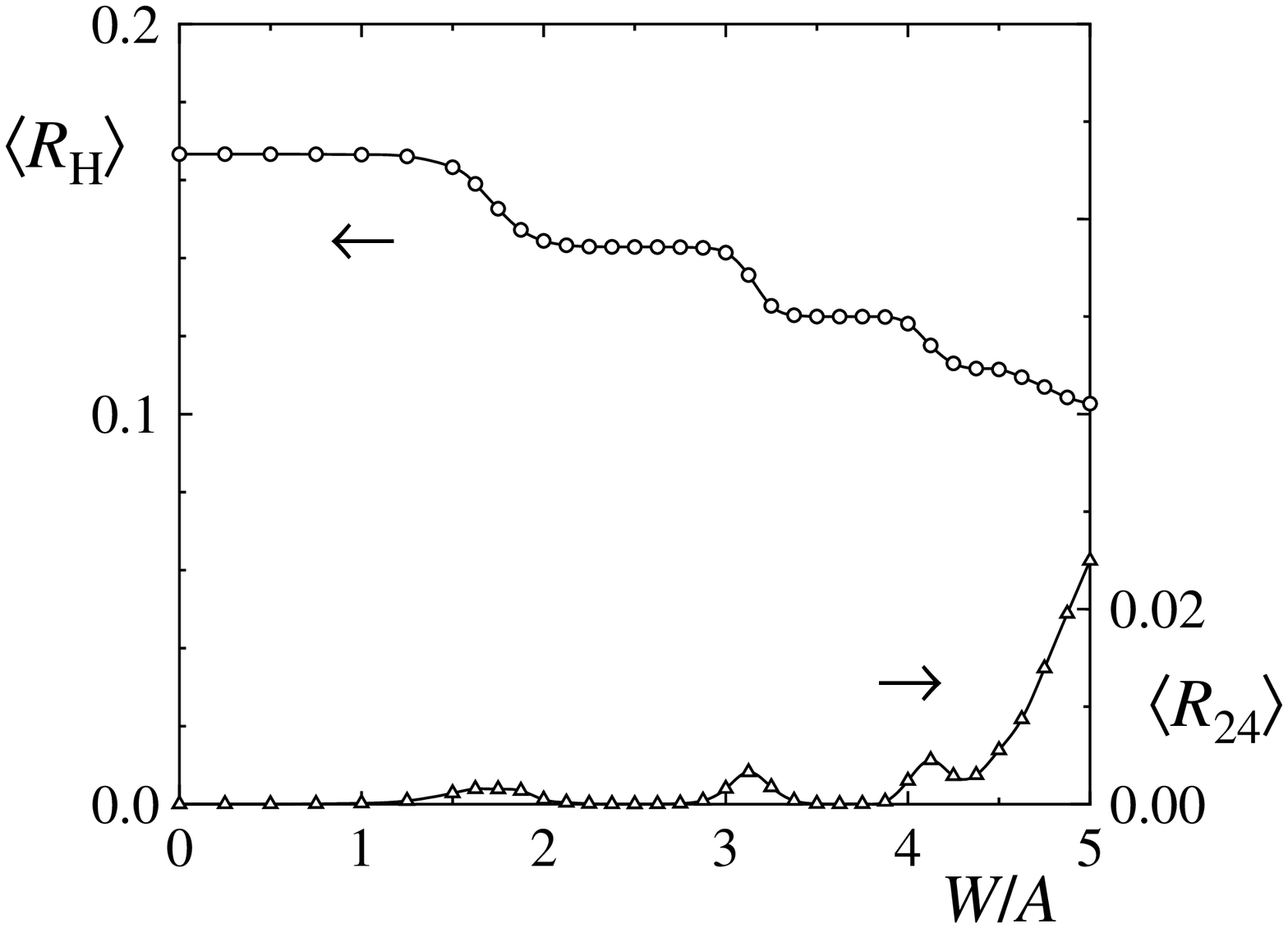}
\end{center}
\caption{
$\langle R_{\rm H} \rangle $ and $\langle R_{24} \rangle$
as functions of $W/A$ in the case of $E_{\rm F}/A = 0$.
Solid lines serve as visual guides.
}
\begin{center}
\includegraphics[height=6.0cm]{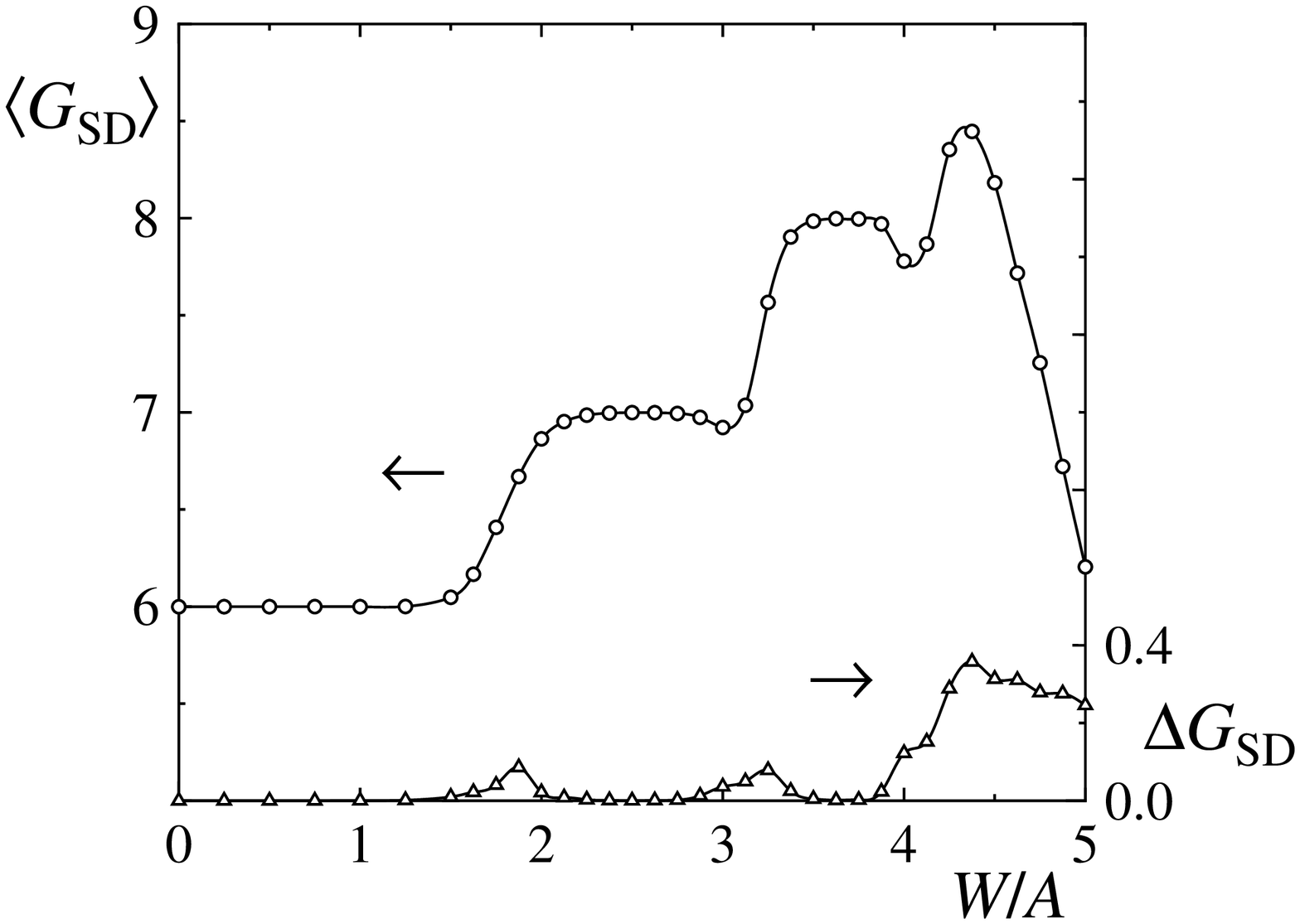}
\end{center}
\caption{
$\langle G_{\rm SD} \rangle $ and $\Delta G_{\rm SD}$
as functions of $W/A$ in the case of $E_{\rm F}/A = 0$.
Solid lines serve as visual guides.
}
\end{figure}
We mainly consider the case where the Fermi energy is
fixed at the Weyl nodes (i.e., $E_{\rm F}/A = 0$).
In Fig.~4, we show $\langle G_{\rm H} \rangle $ and $\Delta G_{\rm H}$
as functions of $W/A$.
A notable feature of $\langle G_{\rm H} \rangle$ is
that it shows successive plateaus, where its value on each plateau
increases from $6$ to $9$ with increasing $W/A$.
Figure~4 also shows that $\Delta G_{\rm H}$ becomes nearly zero on each plateau
except for that corresponding to $G_{\rm H} = 9$ at $W/A \sim 4.4$.
This indicates that $G_{\rm H}$ is quantized to an integer
as long as $W/A \lesssim 4$.
It is natural to consider that
this quantization of $G_{\rm H}$ is caused by chiral edge modes
assisted by a finite-size excitation gap of bulk states at the Weyl nodes.
Near the clean limit of $W/A = 0$, the quantized value of $G_{\rm H}$
is identical to the number of chiral modes (i.e., $6$)
as expected from the analysis given in Sect.~2.
However, the quantized value increases with increasing $W/A$,
indicating that the number of chiral modes also increases.
This should be attributed to the renormalization of a mass term
due to disorder: $2t\cos(k_{0}a)$ in $h_{0}$
is effectively replaced by $2t\cos(k_{0}a) - \delta M$,
where $\delta M \propto (W/A)^{2}$ within a Born approximation.~\cite{groth}
The renormalization of $2t\cos(k_{0}a)$ is rewritten as
\begin{align}
   2t\cos(k_{0}a) - \delta M = 2t\cos\left(({k}_{0}+\delta k)a\right)
\end{align}
with
\begin{align}
   \delta ka \sim \frac{1}{\sin(k_{0}a)}\frac{\delta M}{2t} .
\end{align}
That is, $k_{0}$ effectively increases with increasing $W/A$.
This accounts for the increase in the number of chiral modes
with increasing disorder,
and suggests the possibility to control chiral modes by disorder.
The behavior of $\langle G_{\rm H} \rangle $ as a function of $W/A$
is consistent with that of the Hall conductivity observed
in Refs.~\citen{chen2} and \citen{shapourian}.

The mass renormalization due to disorder has been argued for
2D topological insulators, in which it causes a transition
from nontopological to topological phases.~\cite{groth,yamakage}
In a manner similar to this,
the boundary between the semimetal phase and the diffusive metal phase
is shown to be modified by it in Weyl semimetals.~\cite{chen2,shapourian,LOS}
The above result indicates that, even within a semimetal phase,
the mass renormalization significantly affects the property
of edge excitations.

An incomplete quantization of $G_{\rm H}$
in the region of $W/A \gtrsim 4$ indicates that
chiral edge modes are destabilized owing to disorder.
A plausible explanation is that a finite-size gap at the Weyl nodes
is closed by a strong disorder and hence chiral edge modes located
at one side of the system are coupled with those in the opposite side by
low-energy bulk states,~\cite{LOS}
resulting in the destabilization of chiral edge modes.
As $\Delta G_{\rm H}$ turns to increase near $W/A = 4.3$
without being reduced to zero, we observe that the critical strength
$W_{\rm c}$ of disorder is $W_{\rm c}/A \sim 4.3$ in this case.
This value is consistent with that obtained
by finite-size scaling analysis.~\cite{shapourian}

Figure~5 shows $\langle R_{\rm H} \rangle$ and $\langle R_{24} \rangle$
as functions of $W/A$.
We see that $\langle R_{\rm H} \rangle$ shows a plateau structure,
which is consistent with the behavior of $\langle G_{\rm H} \rangle$,
and that $\langle R_{24} \rangle$ vanishes in the region corresponding to
a plateau of $\langle R_{\rm H} \rangle$ as long as $W/A \lesssim 4$.
Furthermore, $\langle R_{24} \rangle$ tends to monotonically increase
in the region of $W/A \gtrsim 4.3$.
These behaviors support the argument given above.

Figure~6 shows $\langle G_{\rm SD} \rangle$ and $\Delta G_{\rm SD}$
as functions of $W/A$.
In the region of $W/A \lesssim 4$, we see that $\langle G_{\rm SD} \rangle$
shows plateaus, on each of which it takes an integer value,
and that $\Delta G_{\rm SD}$ simultaneously becomes nearly zero.
This indicates that $G_{\rm SD}$ is quantized to an integer.~\cite{yoshimura}
The quantized value of $G_{\rm SD}$ increases
from $6$ to $8$ with increasing disorder,
again indicating the increase in the number of chiral edge modes.
We also see that $\langle G_{\rm SD} \rangle $ starts to decrease
near $W/A \sim 4.3$.
This supports the observation that the critical strength of disorder is
$W_{\rm c}/A \sim 4.3$ in this case.

\begin{figure}[tbp]
\begin{center}
\includegraphics[height=6.0cm]{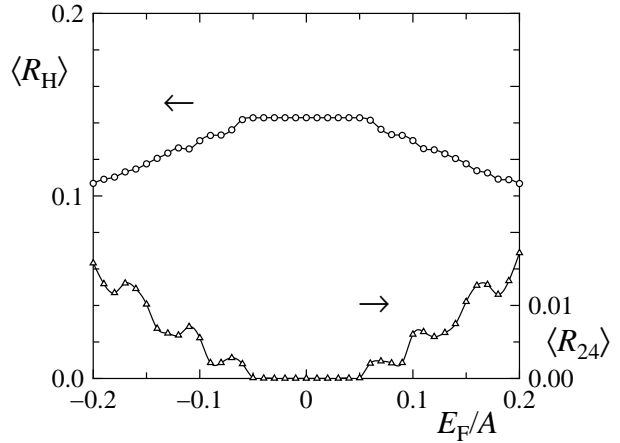}
\end{center}
\caption{
$\langle R_{\rm H} \rangle $ and $\langle R_{24} \rangle $
as functions of $E_{\rm F}/A$ at $W/A = 2.5$.
Solid lines serve as visual guides.
}
\end{figure}
In the remainder of this section, we briefly examine how the quantization of
$R_{\rm H}$ is affected by the variation of $E_{\rm F}/A$
focusing on the case of $W/A = 2.5$,
where $R_{\rm H}$ is precisely quantized to $1/7$ at $E_{\rm F}/A = 0$.
Figure~7 shows $\langle R_{\rm H} \rangle$ and $\langle R_{24} \rangle$
as functions of $E_{\rm F}/A$.
We clearly observe that $\langle R_{\rm H} \rangle$ is nearly equal to $1/7$
inside the region of $|E_{\rm F}|/A \lesssim 0.05$
and decreases with increasing $|E_{\rm F}|/A$ outside the region.
We also observe that $\langle R_{24} \rangle$ is nearly zero
in the region of $|E_{\rm F}|/A \lesssim 0.05$,
where $\langle R_{\rm H} \rangle \approx 1/7$, while it takes a finite value
depending on $E_{\rm F}/A$ outside the region.
This supports the reasoning that
the quantization of $R_{\rm H}$ manifests itself
when the Fermi level is placed within a finite-size gap at the Weyl nodes.
Figure~7 indicates that the finite-size gap $\Delta$ is
on the order of $\Delta/A \sim 0.1$ in this case.

\section{Summary and Discussion}

The disorder effect on Weyl semimetals is studied for a thin slab of
Weyl semimetal hosting Fermi arc surface states
only at its side in the form of chiral edge modes.
Setting the system in a Hall bar geometry,
we numerically calculate the dimensionless Hall conductance
$G_{\rm H}$ in the presence of disorder.
When the Fermi level is located near the Weyl nodes within a finite-size gap,
it is shown that $G_{\rm H}$ is quantized to an integer equal to
the number of chiral modes as long as the strength $W$ of disorder is
smaller than the critical value $W_{\rm c}$,
and that this quantization collapses once $W$ exceeds $W_{\rm c}$.
This indicates that the presence or absence of the quantization
serves as a probe to distinguish
a Weyl semimetal phase from a diffusive anomalous Hall metal phase.
It is also shown that the quantized value of $G_{\rm H}$
increases with increasing $W$ as long as $W < W_{\rm c}$.
This is a direct consequence of the increase in the number of chiral modes
caused by the renormalization of a mass parameter due to disorder.

Finally, note that the quantizations of $G_{\rm H}$ and
$G_{\rm SD}$ shown in Sect.~3 are less precise than
those observed in an ordinary quantum Hall system.
The central reason is that the bulk excitation gap is much smaller than
that in an ordinary quantum Hall system
since it is induced by a finite-size effect.
If the system length $L$ is sufficiently small as in the case
examined in Sect.~3, the bulk gap fully suppresses the backscattering
of chiral edge modes across the bulk despite its smallness.
However, the effect of backscattering gradually becomes stronger
with increasing $L$ and then the precision of the quantizations
eventually decreases owing to this.~\cite{gorbar}

\section*{Acknowledgment}

This work was supported by JSPS KAKENHI Grant Number 15K05130.

\appendix

\section{}

Let us find elementary solutions for $\mib{\varphi}(m)$
in $|\psi(k_{x},k_{z}) \rangle$.
Under the assumption of $\mib{\varphi}(m) = \rho^{m}\mib{v}$,
the eigenvalue equation~(\ref{eq:EE-x}) is reduced to
\begin{align}
     \label{eq:red-eigen_V}
  \left[
    \begin{array}{cccc}
      M(\rho) & -A(\rho) \\
      A(\rho) & -M(\rho)
    \end{array}
  \right] \mib{v}
 & = E_{\perp} \mib{v} ,
\end{align}
where
\begin{align}
  M(\rho)
 & = \Delta(k_{z})+2B-B\left(\rho+\rho^{-1}\right) ,
        \\
  A(\rho)
 & = \frac{1}{2}A\left(\rho-\rho^{-1}\right) .
\end{align}
Equation~(\ref{eq:red-eigen_V}) holds only when
\begin{align}
       \label{eq:eigen-value}
  E_{\perp}^{2}-M(\rho)^{2}+A(\rho)^{2} = 0 .
\end{align}
Let $\mib{\varphi}_{\pm}(m) \equiv \rho_{\pm}^{m} \mib{v}_{\pm}$
be two different elementary solutions of Eq.~(\ref{eq:EE-x}),
in terms of which we can express a general solution as
\begin{align}
  \mib{\varphi}(m)
  = d_{+} \rho_{+}^{m} \mib{v}_{+}
  + d_{-} \rho_{-}^{m} \mib{v}_{-} .
\end{align}
The boundary condition of $\mib{\varphi}(\infty) = {}^{t}(0,0)$ requires
\begin{align}
  |\rho_{\pm}| < 1 .
\end{align}
The other boundary condition of $\mib{\varphi}(0) = {}^{t}(0,0)$ requires
\begin{align}
  \mib{v}_{+} = \mib{v}_{-}
\end{align}
for $\rho_{+} \neq \rho_{-}$ with $d_{-} = - d_{+}$.

Now we consider the case where $\mib{v}_{+} = \mib{v}_{-}$ holds.
It is instructive to rewrite Eq.~(\ref{eq:red-eigen_V}) as
\begin{align}
  \left[
    \begin{array}{cccc}
      \frac{M(\rho_{\pm})-E_{\perp}}{A(\rho_{\pm})}
         & -1 \\
      1 & \frac{-M(\rho_{\pm})-E_{\perp}}{A(\rho_{\pm})}
    \end{array}
  \right] \mib{v}_{\pm}
  =  \mib{0} .
\end{align}
This indicates that $\mib{v}_{+} = \mib{v}_{-}$
is realized for $\rho_{+} \neq \rho_{-}$ only when
\begin{align}
    \frac{M(\rho_{+})-E_{\perp}}{A(\rho_{+})}
    = \frac{M(\rho_{-})-E_{\perp}}{A(\rho_{-})}
\end{align}
and
\begin{align}
    \frac{-M(\rho_{+})-E_{\perp}}{A(\rho_{+})}
    = \frac{-M(\rho_{-})-E_{\perp}}{A(\rho_{-})}
\end{align}
simultaneously hold.~\cite{imura2}
These equations require that
$E_{\perp} = 0$, with which Eq.~(\ref{eq:eigen-value}) yields
\begin{align}
     \label{eq:M-A}
   M(\rho_{\pm}) = \pm A(\rho_{\pm}) .
\end{align}
As shown later, solutions with $|\rho_{\pm}| < 1$ are always obtained
in the case of $M(\rho_{\pm}) = A(\rho_{\pm})$ under the condition of
$4B > -\Delta(k_{z}) > 0$ with $A > 0$ and $B > 0$.
This immediately yields
\begin{align}
     \label{eq:rho-pm}
  \rho_{\pm}
  = \frac{\Delta(k_{z})+2B
          \pm\sqrt{(\Delta(k_{z})+2B)^{2}-4B^{2}
                   +A^{2}}}
         {2\left(B+\frac{A}{2}\right)}
\end{align}
and
\begin{align}
  \mib{v}
  = \frac{1}{\sqrt{2}}
    \left[ \begin{array}{c}
             1 \\
             1
           \end{array}
    \right] ,
\end{align}
where $\mib{v} \equiv \mib{v}_{+}=\mib{v}_{-}$.
The wave function $\mib{\varphi}(m)$ is finally expressed as
\begin{align}
          \label{eq:varphi_AP}
   \mib{\varphi}(m)
   = \mathcal{C} \left(\rho_{+}^{m}-\rho_{-}^{m}\right)\mib{v} ,
\end{align}
where $\mathcal{C}$ is a constant to be determined
by the normalization condition of
$\sum_{m=1}^{\infty}|\mathcal{C}\left(\rho_{+}^{m}-\rho_{-}^{m}\right)|^{2}=1$.
Equation~(\ref{eq:varphi_AP}) is equivalent to Eq.(\ref{eq:varphi_0}).

Now we turn to Eq.~(\ref{eq:M-A})
and show that solutions with $|\rho_{\pm}| < 1$ are obtained only
in the case of $M(\rho_{\pm})=A(\rho_{\pm})$ under the condition of
$4B > -\Delta(k_{z}) > 0$ with $A > 0$ and $B > 0$.
To do so, let us examine the two cases of $M(\rho_{\pm})=A(\rho_{\pm})$
and $M(\rho_{\pm})=-A(\rho_{\pm})$.
In the first case, $\rho_{\pm}$ is obtained as
\begin{align}
     \label{eq:1pm}
  \rho_{1\pm}
  = \frac{\Delta(k_{z})+2B
          \pm\sqrt{D}}
         {2\left(B+\frac{A}{2}\right)} ,
\end{align}
while
\begin{align}
     \label{eq:2pm}
  \rho_{2\pm}
  = \frac{\Delta(k_{z})+2B
          \pm\sqrt{D}}
         {2\left(B-\frac{A}{2}\right)}
\end{align}
in the second case, where
\begin{align}
   D 
   \equiv (\Delta(k_{z})+2B)^{2}-4B^{2}+A^{2} .
\end{align}
We show below that $|\rho_{1\pm}| < 1$ always holds
while $|\rho_{2\pm}| < 1$ never holds.
That is, the appropriate solutions are obtained in the case of
$M(\rho_{\pm})=A(\rho_{\pm})$.
We separately consider the cases of $D < 0$ and $D > 0$ below.

\subsection{Case of $D < 0$}

In this case, Eqs.~(\ref{eq:1pm}) and (\ref{eq:2pm}) are rewritten as
\begin{align}
  \rho_{1\pm}
  = \frac{\Delta(k_{z})+2B
          \pm i \sqrt{-D}}
         {2\left(B+\frac{A}{2}\right)}
  = \rho_{2\pm}^{-1} .
\end{align}
This immediately yields
\begin{align}
  |\rho_{1\pm}|
  = \sqrt{\frac{B-\frac{A}{2}}{B+\frac{A}{2}}}
  < 1 <
    \sqrt{\frac{B+\frac{A}{2}}{B-\frac{A}{2}}}
  = |\rho_{2\pm}| .
\end{align}
Note that $D < 0$ is satisfied under the condition of
\begin{align}
  2B+\sqrt{4B^{2}-A^{2}} > -\Delta(k_{z}) > 2B-\sqrt{4B^{2}-A^{2}}
\end{align}
with $2B > A > 0$.

\subsection{Case of $D > 0$}

In this case, we can show from Eqs.~(\ref{eq:1pm}) and (\ref{eq:2pm})
that $\rho_{1\pm}\rho_{2\mp} = 1$,
and that $|\rho_{2+}| > |\rho_{1+}|$ and $|\rho_{2-}| > |\rho_{1-}|$
since $A > 0$ and $B > 0$ are assumed.
Let us separately treat the two cases of $\Delta(k_{z})+2B > 0$
and $\Delta(k_{z})+2B < 0$.

If $\Delta(k_{z})+2B > 0$,
we find that $|\rho_{2+}| > |\rho_{2-}|$ from Eq.~(\ref{eq:2pm}).
The combination of this with $|\rho_{2-}| > |\rho_{1-}|$ yields
$|\rho_{2+}| > |\rho_{1-}|$, indicating that
\begin{align}
  |\rho_{2+}| > 1 > |\rho_{1-}|
\end{align}
since $\rho_{1-}\rho_{2+} = 1$.
Thus, $|\rho_{2\pm}| < 1$ never holds.
Here, we also find from Eq.~(\ref{eq:1pm}) with $\Delta(k_{z})+2B > 0$
that $\rho_{1+} > |\rho_{1-}|$.
The above argument indicates that $|\rho_{1\pm}| < 1$ when
\begin{align}
  1 > \rho_{1+} .
\end{align}
This holds when $2B > -\Delta(k_{z}) > 0$.
Let us examine the compatibility of this condition and $D > 0$.
If $A > 2B > 0$, we find that $D > 0$ is always satisfied
under the condition of
\begin{align}
   2B > -\Delta(k_{z}) > 0 .
\end{align}
If $2B > A > 0$, we find that they are simultaneously satisfied
under the condition of
\begin{align}
  2B-\sqrt{4B^{2}-A^{2}} > -\Delta(k_{z}) > 0 .
\end{align}

If $\Delta(k_{z})+2B < 0$,
we find that $|\rho_{2-}| > |\rho_{2+}|$ from Eq.~(\ref{eq:2pm}).
The combination of this with $|\rho_{2+}| > |\rho_{1+}|$ yields
$|\rho_{2-}| > |\rho_{1+}|$, indicating that
\begin{align}
  |\rho_{2-}| > 1 > |\rho_{1+}|
\end{align}
since $\rho_{1+}\rho_{2-} = 1$.
Thus, $|\rho_{2\pm}| < 1$ never holds.
Here, we also find from Eq.~(\ref{eq:1pm}) with $\Delta(k_{z})+2B < 0$
that $|\rho_{1-}| > |\rho_{1+}|$.
The above argument indicates that $|\rho_{1\pm}| < 1$ when
\begin{align}
  1 > |\rho_{1-}| .
\end{align}
This holds when $4B > -\Delta(k_{z}) > 2B$.
Let us examine the compatibility of this condition and $D > 0$.
If $A > 2B > 0$, we find that $D > 0$ is always satisfied
under the condition of
\begin{align}
   4B > -\Delta(k_{z}) > 2B .
\end{align}
If $2B > A > 0$, we find that they are simultaneously satisfied
under the condition of
\begin{align}
  4B > -\Delta(k_{z}) > 2B+\sqrt{4B^{2}-A^{2}} .
\end{align}

Combining the results of all the cases, we conclude that $|\rho_{\pm}| < 1$
is realized in the case of $M(\rho_{\pm})=A(\rho_{\pm})$
under the condition of $4B > -\Delta(k_{z}) > 0$ with $A > 0$ and $B > 0$.
Contrastingly, we find that $|\rho_{\pm}| < 1$ does not hold
in the case of $M(\rho_{\pm})=-A(\rho_{\pm})$.
Note that, as $2B > t[1-\cos(k_{0}a)]$ is assumed in Eq.~(\ref{eq:assump_B-t}), the condition of $4B > -\Delta(k_{z}) > 0$ is simply equivalent to
\begin{align}
   k_{0} > k_{z} > - k_{0} .
\end{align}

\end{document}